\documentclass[12pt]{article}
\usepackage{graphicx}

      \usepackage[cp1251]{inputenc} 

\begin{document}
 \noindent {\footnotesize\it\small IV Finnish-Russian Radio Astronomy
 Symposium, 15-18 October 2012, Lammi Biological Station, Finland}

 \noindent
 \begin{tabular}{llllllllllllllllllllllllllllllllllllllllllll}
 & & & & & & & & & & & & & & & & & & & & & & & & & & & & & & & & & & & & \\\hline\hline
 \end{tabular}

 \vskip 1.0cm
 \begin{center} {\Large\bf Phase Retrieval Problem.\\
  Application to VLBI Mapping of Active Galactic Nuclei}
  \end{center}

 \bigskip
 \centerline {A.T. Bajkova}
 \medskip
\centerline{\small\it
 Pulkovo Astronomical Observatory, Russian Academy of Sciences,
 St.-Petersburg, Russia
 }

 \bigskip
{\bf Abstract}---The well-known phase problem which means image
reconstruction from only spectrum magnitude without using any
spectrum phase information is considered basically in application
to VLBI mapping of compact extragalactic radio sources (active
galactic nuclei). The method proposed for phaseless mapping is
based on the reconstruction of the spectrum magnitude on the
entire $UV$ plane from the measured visibility magnitude on a
limited set of points and the reconstruction of the sought-for
image of the source by Fienup’s error-reduction iterative
algorithm from the spectrum magnitude reconstructed at the first
stage. It is shown that the technique used ensures unique solution
(within a class of equivalent functions) for AGNs with typical
structure morphology "bright core + week jet". The method proposed
can be used, for example, for imaging with ultra-high resolution
using a space--ground radio interferometer with a space antenna in
a very high orbit ("RadioAstron"). In this case, a multi-element
interferometer essentially degenerates into a two-element
interferometer and the degeneracy of the close-phase relations
prevents the use of standard methods for hybrid mapping and
self-calibration. The capabilities and restrictions of the method
are demonstrated on a number of model experiments. For a few
selected AGNs the images are obtained from VLBA observations.

\section{Introduction to phase retrieval problem}

In this paper we consider the problem of image reconstruction from
only spectral magnitude when spectral phase of an object is
totally unknown.

Why this problem is important? In general case, as was shown by
many authors, phase information is more important than magnitude
information, because phase information is responsible for
positions of details of image. Importance of spectral phase is
clear from Fig.1. In Fig.1(a) the picture of a man is shown. The
image (b), reconstructed using original Fourier spectrum phase
information of the model image and uniform spectral magnitude
allows us to recognize the features of the portrait. The image (c)
reconstructed from original spectral magnitude and zero-valued
spectral phase does not give anything common with original
portrait (a).

Let us formulate the phase problem in discrete form. Let the
discretization in spatial and spectrum domains of an object be
fulfilled in accordance with Kotel'nikov-Shannon theorem and size
of two-dimensional maps be equal to $N\times N$ samples.

Spectrum of an object is calculated as discrete two-dimensional
Fourier transform of two-dimensional sequence $x_{ml},
m,l=0,...,N$ as follows:
\begin{figure}[t]
{\begin{center}
 \includegraphics*[width=155mm]{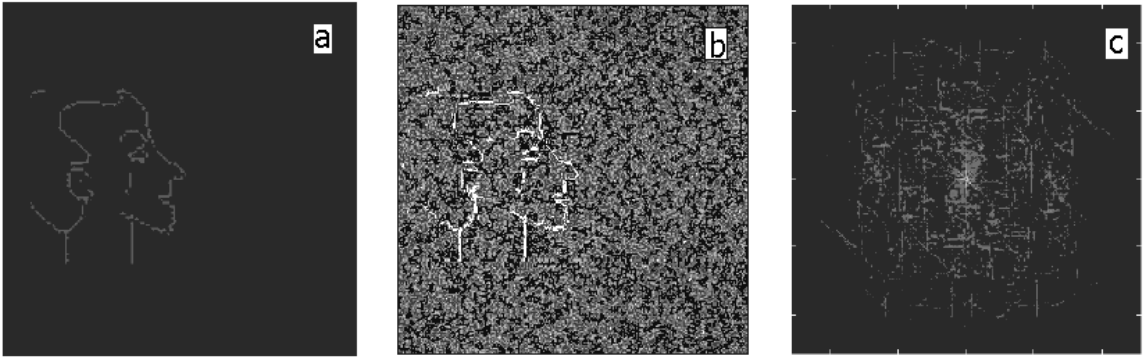}
 \caption
 {
Importance of phase: (a)model: all nonzero points=1, (b)
corresponding image with original spectral phase and uniform
spectral magnitude, (c)corresponding image with original spectral
magnitude and zero-valued spectral phase}

\end{center}}
\end{figure}

\begin{equation}
X_{nk}=\frac{1}{N}\sum^{N-1}_{m=0}\sum^{N-1}_{l=0}x_{ml}\exp(-\frac{j2\pi}{n}(nm+kl))=A_{nk}+jB_{nk}=M_{nk}\exp(j\Phi_{nk}),
\end{equation}
where $n,k=1,...,N$, $A_{nk}, B_{nk}, M_{nk}$ and $\Phi_{nk}$ are
real part, imaginary part, magnitude and phase of spectrum
correspondingly.

We consider the case of image formation systems when data are
delivered not directly on an object but on its Fourier spectrum.
If data are complete, i.e. we have both magnitude and phase of
spectrum, then the image sought-for can be obtained by simple
inverse Fourier transform of data:

\begin{equation}
x_{ml}=\frac{1}{N}\sum^{N-1}_{n=0}\sum^{N-1}_{k=0}M_{nk}\exp(j\Phi_{nk})\exp(\frac{j2\pi}{n}(nm+kl)),
\end{equation}

But if the data are not complete, i.e. we have information only
about magnitude or only about phase of the spectrum, the problem
of retrieval of missing information arises, because only presence
of both components gives correct representation of an object. Here
we consider a typical for many fields problem of phase or,
equivalently, image $x_{ml}$ retrieval from its Fourier spectrum
magnitude $M_{nk}$. Such a problem is called "phase problem". The
most significant works devoted to phase problem are published by
Gerchberg and Saxton (1972), Fienup (1978, 10982, 1983),
Fienup,Crimmins and Holsztynski (1982), Fienup et al (2006, 2009),
Huang, Bruck and Sodin (1979), Hayes (1982), Oppenheim and Lim
(1981), Dainty and Fiddy (1984),Sanz and Huang (1983).

\section{The uniqueness of the solution}

In the most general formulation where constraints are imposed only
on the spectrum magnitude, the problem of reconstructing the
function has an infinite set of solutions. Indeed, any function
that has a given spectrum magnitude and an arbitrary spectral
phase satisfies these constraints, and, if at least one solution
is known, the other can be obtained by convolving this solution
with a function that has an arbitrary phase and a spectrum
magnitude equal to unity at all frequencies.

However, a significant narrowing of the set of solutions is
possible for certain constraints imposed on the function being
reconstructed in the spatial domain (Stark 1987). One of these is
the constraint imposed on the spatial extent of an object; i.e.,
the sought-for function must have a finite carrier. Another severe
constraint in the spatial domain is the requirement of real
valuedness  and nonnegativity. Below, in solving the phase
retrieval problem, we assume that the sought-for function
satisfies these constraints; i.e., it is real and nonnegative and
has a finite extent. Further we always will assume that sought-for
solution to $x_{ml}$ is non-negative and has rectangular support
$R(m,l): m=M_1,...,M_2, l=L1,...,L_2$, where $|M_2-M_1|\le N/2,
|L_2-L_1|\le N/2$ in order to compute spectrum modulus without
aliasing.

The finite extent of an object (the finiteness of the function)
ensures that the Fourier spectrum is analytic in accordance with
the Wiener-Paley theorem (Khurgin and Yakovlev 1971). As a result,
this spectrum can be reconstructed from the known part of it,
which is used to reconstruct images from the visibility function
measured on a limited set of points in the $UV$ plane.

If we determine the class of equivalent functions to within a
linear shift and reversal of the argument (rotation through
$180^o$), then all of the functions that belong to this class have
the same spectrum magnitude. The solution of the phase retrieval
problem is assumed to be unique if it was determined to within the
class of equivalent functions.

In the case of one-dimensional functions, even these severe
constraints on finiteness and nonnegativity do not guarantee a
unique reconstruction from the spectrum magnitude. As the
dimensionality of the function increases ($n\ge 2$), a unique (to
within the class of equivalent functions) solution becomes
possible, except for the degenerate cases defined on the set of
measure zero (Bruck and Sodin 1979; Hayes 1982). This follows from
the qualitative difference between the properties of the $z$
-transformations of one-dimensional and multidimensional
sequences.

For a unique solution to exist,the $z$-transformation must be
irreducible, which is not achievable in principle in the
one-dimensional case and almost always holds in the
multidimensional case. Since we deal with two-dimensional images
in VLBI, we assume that the solution of the phase retrieval
problem exists and is unique. However, the existence of a unique
solution does not yet guarantee that the retrieval algorithms
converge. The papers by Gerchberg and Saxton (1972)and Fienup
(1978) are of greatest importance in developing the theory and
algorithms of solving the phase retrieval problem.

\section{Fienup's error-reduction phase retrieval algorithm}

Here we consider one of the most efficient algorithm for image
reconstruction from the spectrum magnitude of an object, proposed
by Fienup, so called error-reduction algorithm.

Recall that Fienup’s error-reduction algorithm (Fig.2) is an
iterative process of the passage from the spatial domain of an
object to the spatial frequency domain and back (using the direct
and inverse Fourier transforms) in an effort to use the input
constraints on the spectrum magnitude of the object in the
frequency domain and the constraints on its nonnegativity and
finite extent in the spatial domain. The finite carrier within
which the source is expected to be localized should be specified
in the form of a rectangle. Fienup et al.(1982) estimated the size
of the carrier from the autocorrelation function equal to the
inverse Fourier transform of the square of the spectrum magnitude
for the object. Very important advantages of Fienup’s algorithm
are its high stability against noise (Sanz and Huang 1983)
(stability means that a small change in input data causes the
solution to change only slightly) and high speed (due to the
application of fast Fourier transform algorithms) compared to
other algorithms.

In practice, the error-reduction algorithm usually decreases the
error rapidly for the first few iterations but much more slowly
for later iterations. In all our further experiments we used 100
iterations. But this algorithm does not guarantee conversion to
the true solution having stagnation problem and depending on the
initial approximation. The problem of convergence is hard
particular in case of complicated images. One of examples of
stagnation  is shown in Fig.3, from which we can see that phase
retrieval algorithm did not give true solution. Note that in all
experiments dealing with the Fienup's error-reduction algorithm we
used $\delta$-function as an initial approximation.

\begin{figure}[t]
{\begin{center}
 \includegraphics*[width=100mm]{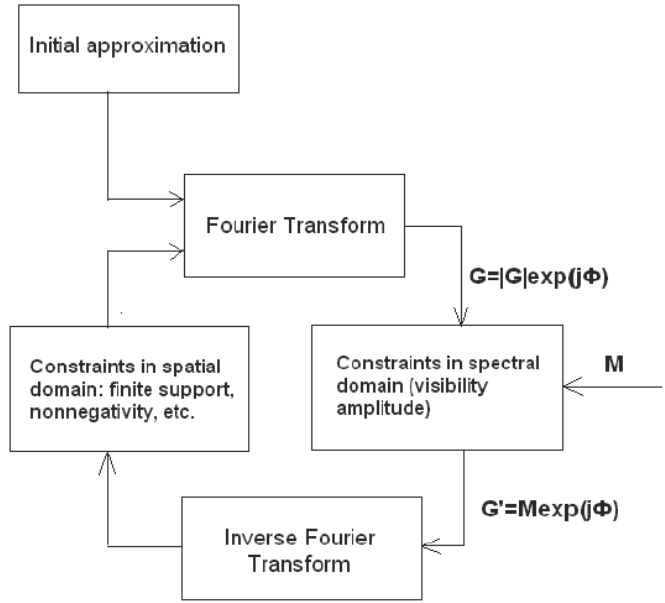}
 \caption{
Fienup's error-reduction phase retrieval algorithm}

\end{center}}
\end{figure}

\section{The role of a reference point}

In this section we demonstrate the essential role of a reference
point, which may be introduced into the image. As it is seen from
Fig.3, in case of absence of some dominated point in the object
the phase retrieval algorithm can fail. Fig.4 shows results of
phase retrieval for different values of the reference point which
is located at the edge of the image as it is shown in Fig.3 (a).
We can see from Fig.4 that the increase of the reference point
value leads to improving the reconstruction quality. With increase
of the reference point value the structure of the object becomes
more and more apparent in the images shown in the left column of
Fig.4, which are obtained using inverse Fourier transform from
original spectral magnitude and zero-valued spectral phase. We can
conclude that spectral magnitude contains more information on
spectral phase for images with higher reference point values.
Therefore Fienup's error-reduction algorithm for such images works
much better ensuring unique true solution. Location of a reference
point may be arbitrary: outside, at the edge or inside of image.
In Fig.5 the case of location of the reference point inside of
image is shown.

\begin{figure}[t]
{\begin{center}
 \includegraphics*[width=155mm]{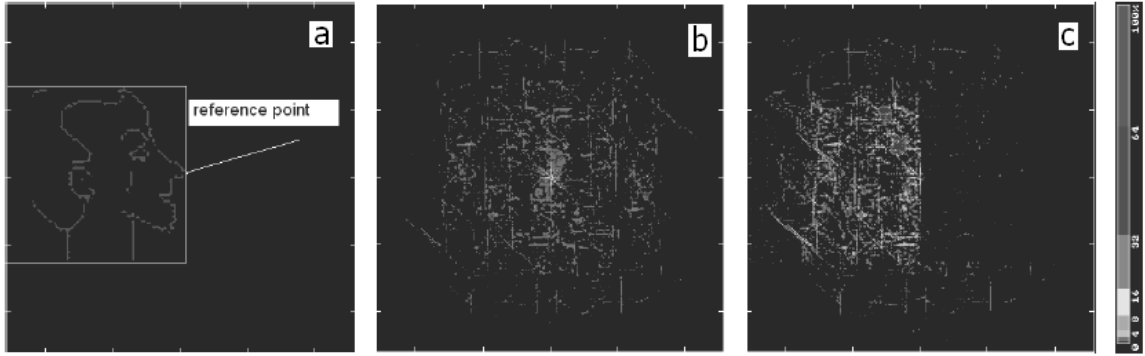}
 \caption{
Image reconstruction result for an object without dominating
reference point: (a) model: all nonzero points=1, reference point
is located at the edge of object, reference point value=1,
rectangular boundaries of the object are given in white color, (b)
corresponding image with original spectral amplitude and
zero-valued spectral phase, (c) Fienup’s algorithm output image}

\end{center}}
\end{figure}

\begin{figure}[p]
{\begin{center}
 \includegraphics*[width=80mm]{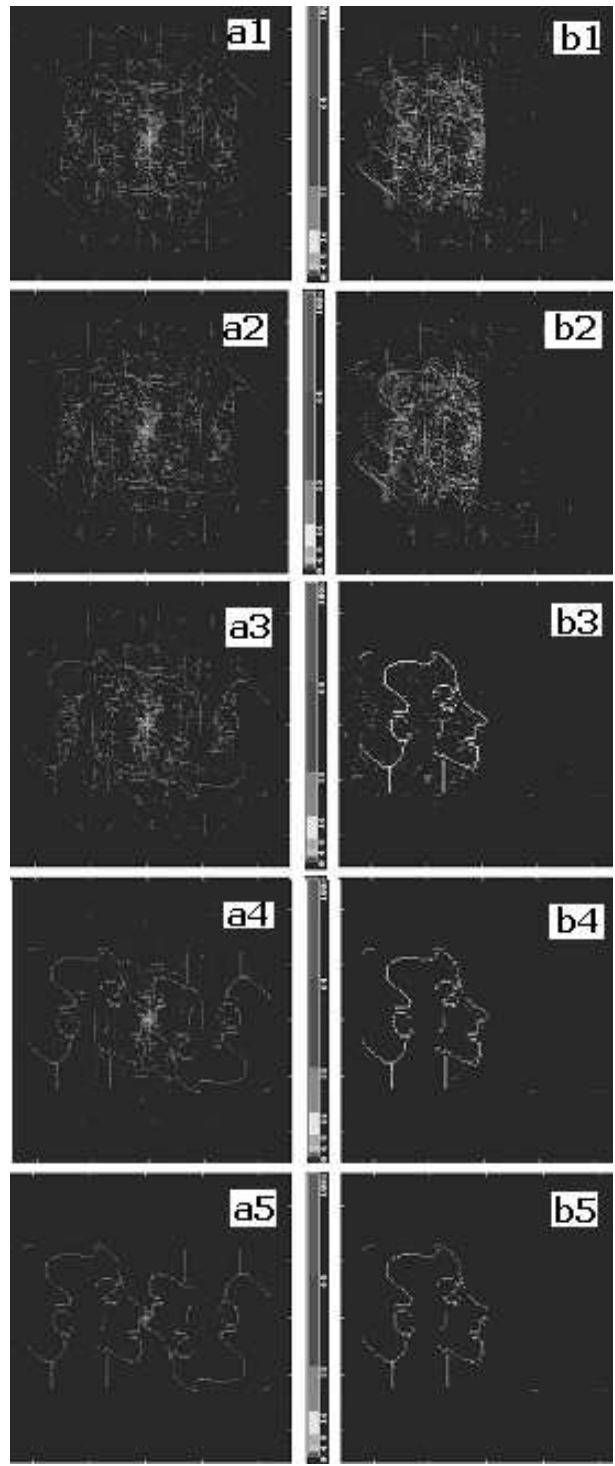}
 \caption{
Reconstruction quality for images with different reference point
values (the reference point is located at the edge of the image):
left column (a):images with original spectrum magnitude and
zero-valued spectrum phase, right column (b): Fienup's algorithm
output images; (a1,b1): reference point value=2, (a2,b2):
reference point value=5, (a3,b3): reference point value=10,
(a4,b4): reference point value=20, (a5,b5): reference point
value=50}

\end{center}}
\end{figure}

\begin{figure}[t]
{\begin{center}
 \includegraphics*[width=150mm]{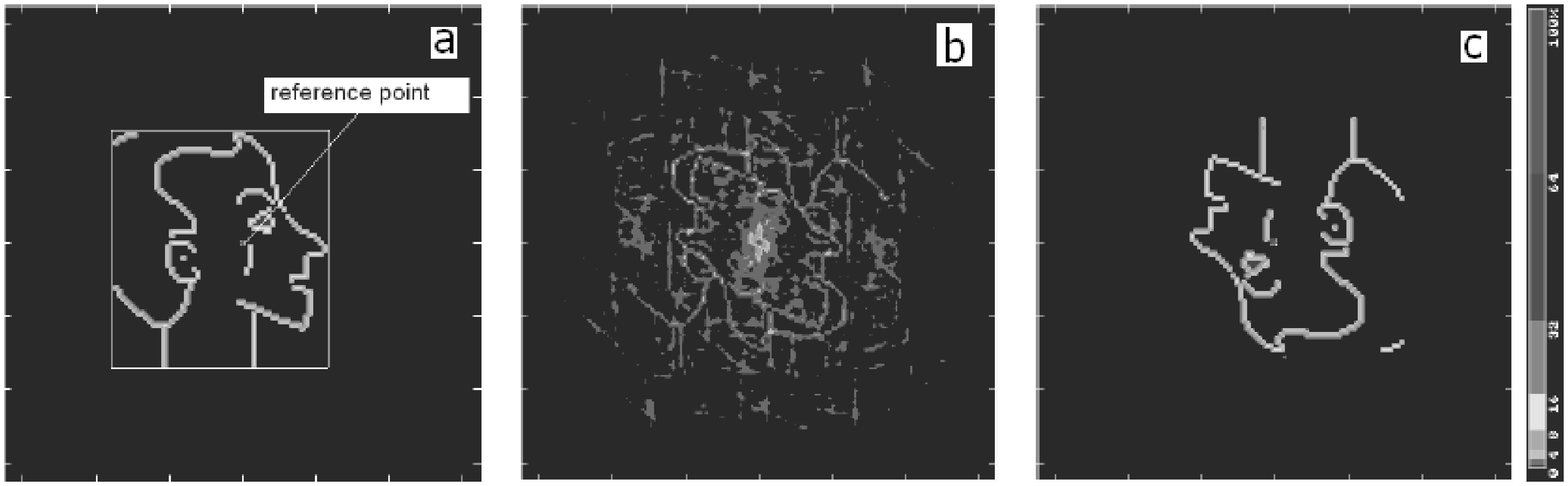}
 \caption{
A reference point is located inside of the image: (a) model: all
nonzero points=1, reference point value=20, (b)corresponding image
with original spectral amplitude and zero-valued spectral phase,
(c) Fienup’s algorithm output image}

\end{center}}
\end{figure}

\section{Application of Fienup's error-reduction algorithm to imaging AGNs}

Above we established that presence in the image of a strong
dominated point leads to unique solution of phase retrieval
problem. Most of extragalactic radio sources (active galactic
nuclei) characterizing by structure morphology "bright compact
core + week jet" meet this requirement. Results of modelling of
phase retrieval problem for a few sources with different positions
and amplitudes of gaussian components are shown in Fig.6. We can
see that images obtained from original spectral magnitude and
zero-valued spectral phase (column(c)) contain information about
real structure (equivalently spectral phase) of an object. Output
images of Fienup's phase retrieval algorithm visually do not
differ from model images shown in column (a). In all considered
cases compact core of sources (at the center of maps) strongly
dominate over other components. In case of less dominating of core
the result of reconstruction is much worse, as it is demonstrated
in Fig.7. So, requirement of strong dominating of compact core
over more week components of the source is very important for
reliable phase retrieval (equivalently image reconstruction) from
only spectral magnitude. Fortunately, practically all AGNs meet
this requirement.

\begin{figure}[p]
{\begin{center}
 \includegraphics*[width=140mm]{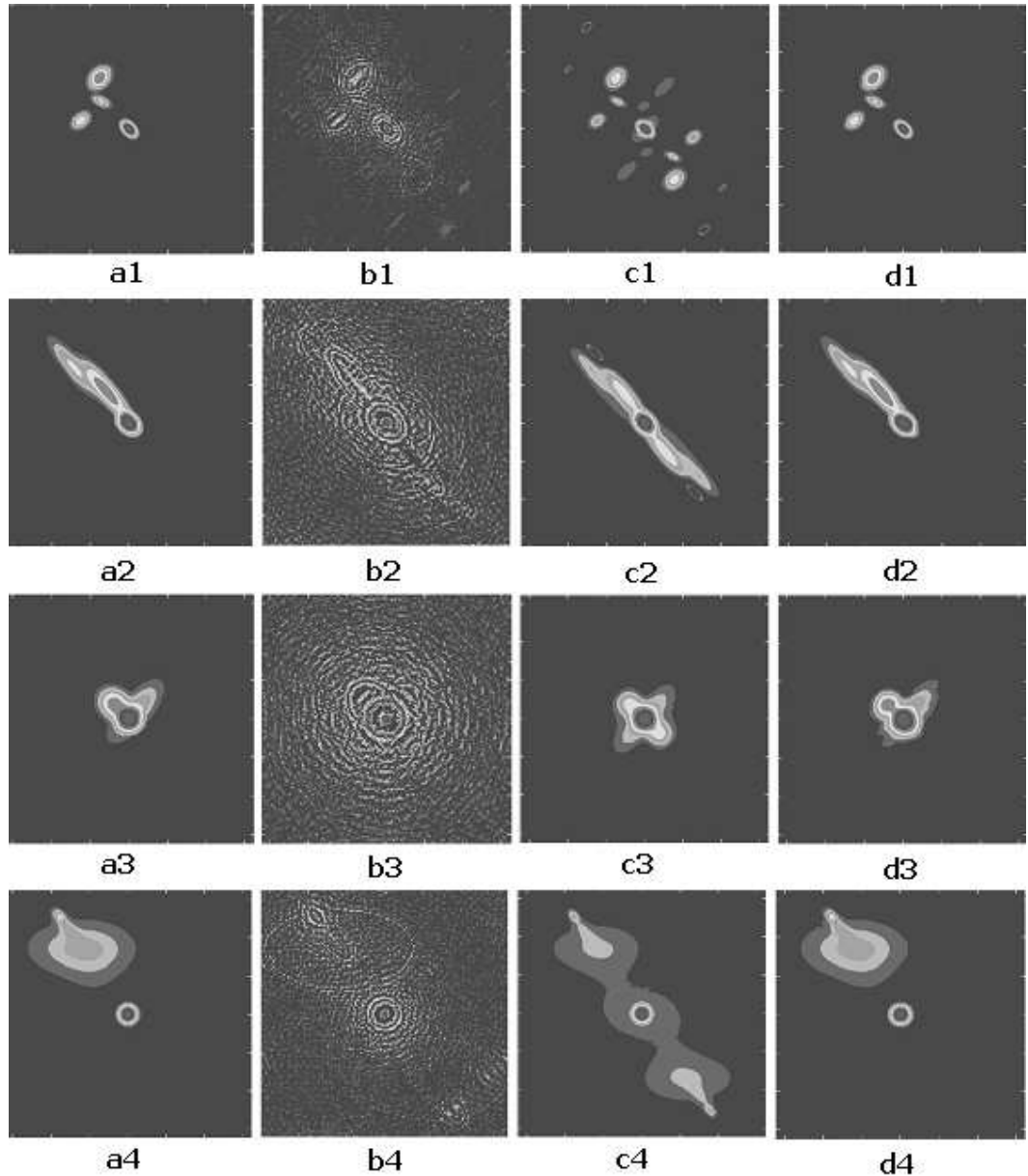}
 \caption{
Modelling of the phase retrieval procedure for compact sources
with different morphology: columns: (a): models, (b): images with
original spectral phase and uniform spectral magnitude, c): images
with original spectral magnitude and zero-valued spectral phase,
(d): Fienup’s algorithm output images; rows: (1) source-1, (2)
source-2, (3) source-3, (4) source-4}

\end{center}}
\end{figure}

\section{"Phaseless" mapping in VLBI }

The algorithms for image reconstruction from the spectral
magnitude of an object, the most efficient of which are Fienup’s
algorithm and its various modifications, require knowledge of the
entire input two-dimensional sequence of sampled points. In VLBI,
however, the visibility function is measured only on a limited set
of points in the $UV$ plane, revealing large unfilled areas and a
diffraction-limited constraint. Therefore, prior reconstruction of
the visibility magnitude on the entire $UV$ plane from a limited
data set is required to successfully use existing reconstruction
algorithms like Fienup’s algorithm.

Thus,the suggested phaseless aperture synthesis method consists of
the following steps:(1)prior reconstruction of the visibility
magnitude (the object ’s spectrum)on the entire $UV$ plane, and
(2)reconstruction of the sought-for image using Fienup’s algorithm
or its modifications (Fienup 1978,1982; Bajkova 2004,2005) from
the spectrum magnitude reconstructed at the first step of the
method. The first step is performed through the reconstruction of
an intermediate image that satisfies the measured visibility
magnitude and a zero phase. Clearly, the intermediate image is
symmetric relative to the phase center of the map. The Fourier
transform of the image obtained yields the spectrum magnitude of
the source extrapolated to the region of the $UV$ plane where
there are no measurements.

The intermediate image can be reconstructed from the visibility
magnitude by the standard method of analytic continuation of the
spectrum using the non-linear CLEAN deconvolution procedures or
the maximum entropy method (MEM)(Cornwell et al.1999). Therefore,
an important requirement for the visibility magnitude is its
analyticity. As we noted in the previous section, this condition
is satisfied for most of the compact extragalactic radio sources
if the flux from the central compact component dominates over the
flux from the remaining fainter components.

Note an important point concerning the use of the maximum entropy
method. In contrast to the CLEAN method: the solution based on the
standard MEM is strictly positive, while the sought-for
intermediate image with a zero spectral phase is generally
alternating, taking on both positive and negative values.

\begin{figure}[t]
{\begin{center}
 \includegraphics*[width=140mm]{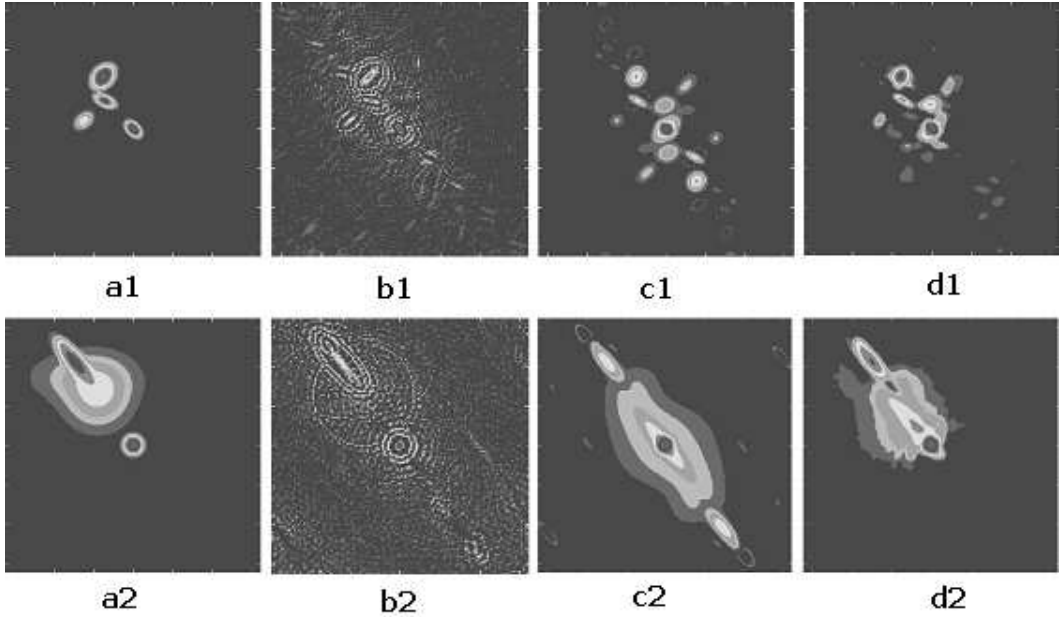}
 \caption{
Modelling of the phase retrieval procedure for AGNs with two
comparable bright components: columns: (a): models, (b): images
with original spectral phase and uniform spectral magnitude, c):
images with original spectral magnitude and zero-valued spectrum
phase, (d): Fienup’s algorithm output images; rows: (1) source-5,
(2) source-6}

\end{center}}
\end{figure}

For reconstruction of images of real sign-variable and complex
functions the generalized maximum entropy method has been
developed by Baikova (1992) (see also Frieden, Bajkova (1994)).
This method is implemented in the Pulkovo VLBI data reduction
software package VLBImager.

Note also that because the GMEM is designed for the reconstruction
of sign-variable functions, it allows one to obtain unbiased
solutions. The bias of the solution is one of the problems of the
conventional MEM (Cornwell et al. 1999), which may lead to a
substantial nonlinear distortion of the final image if the data
contain errors.

For the GMEM, the Shannon-entropy functional has the form
\begin{eqnarray}
E(\alpha) = -\int(x^p(t) ln(\alpha x^p(t)) + x^n(t) ln(\alpha
x^n(t))) dt,\\
x^p(t)> 0, x^n(t) > 0,
\end{eqnarray}
where $x^p(t)$ and $x^n(t)$ are the positive and negative
components of the sought-for image $x(t)$ (Fig.8), i.e. the
equation $x(t) = x^p(t) - x^n(t)$ holds. $\alpha$ > 0 is a
parameter responsible for the accuracy of the separation of the
negative and positive components of the solution $x(t)$, and
therefore critical for the resulting image fidelity.

It is easy to show (Baikova 1992) that solutions for $x^p(t)$ and
$x^n(t)$ obtained with the Lagrange optimization method are
connected by the expression $x^p(t)\times x^n(t) = exp(-2 - 2
ln\alpha) = K(\alpha)$, which depends only on the parameter
$\alpha$. This parameter is responsible for dividing the positive
and negative parts of the solution: the larger $\alpha$ is, the
more accurate is the discrimination. On the other hand, the value
of $\alpha$ is constrained by computational limitations.

\begin{figure}[t]
{\begin{center}
 \includegraphics*[width=86mm]{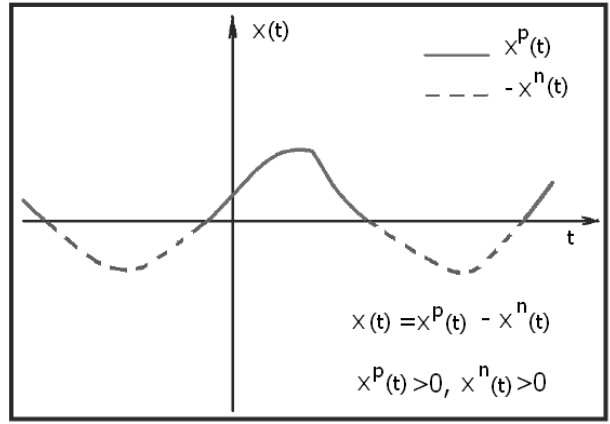}
 \caption{
Form of Generalized MEM solution}

\end{center}}
\end{figure}

Results of modelling of "phaseless" VLBI mapping using the GMEM
and Fienup's phase retrieval algorithm for three model sources are
shown in Fig.9. As can be seen, the results obtained are of quite
satisfactory quality. It is necessary to note, that solution of
the phase problem in case of incomplete magnitude data
significantly depend on accuracy of GMEM intermediate images.
Result of the GMEM reconstruction depends on quality of data. We
investigated quality of imaging for typical level of additive
noise and different level of amplitude errors in visibility data.
Results are shown in Fig.10. We can see that the reconstruction
method is very stable to errors in data and even in the case of
large amplitude errors we have output images which adequately
reflect basic features of the source.

\begin{figure}[p]
{\begin{center}
 \includegraphics*[width=140mm]{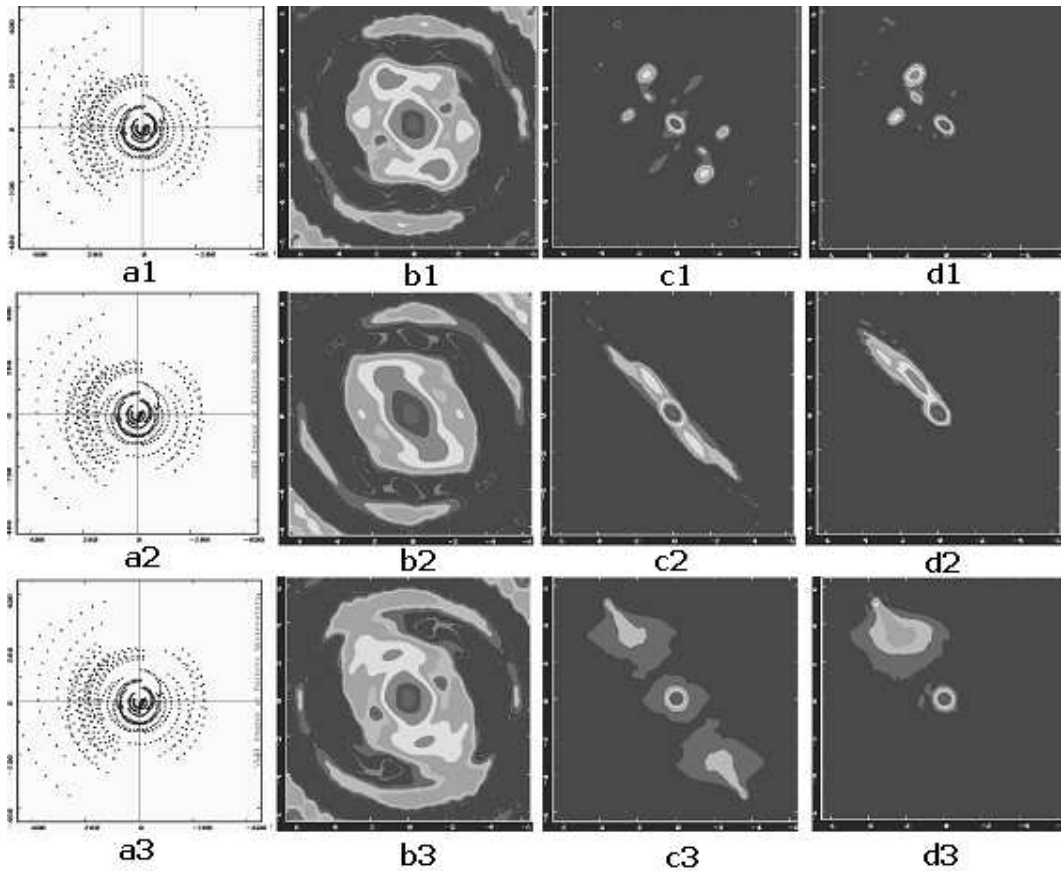}
 \caption{
Modelling of "phaseless" mapping in VLBI: columns: (a): $UV$
coverage, (b): "dirty" images , c): reconstructed intermediate
GMEM images with zero-valued spectral phase, (d): Fienup’s
algorithm output images; rows: (1) for source-1, (2) for source-2,
(3) for source-4}

\end{center}}
\end{figure}

\begin{figure}[t]
{\begin{center}
 \includegraphics*[width=140mm]{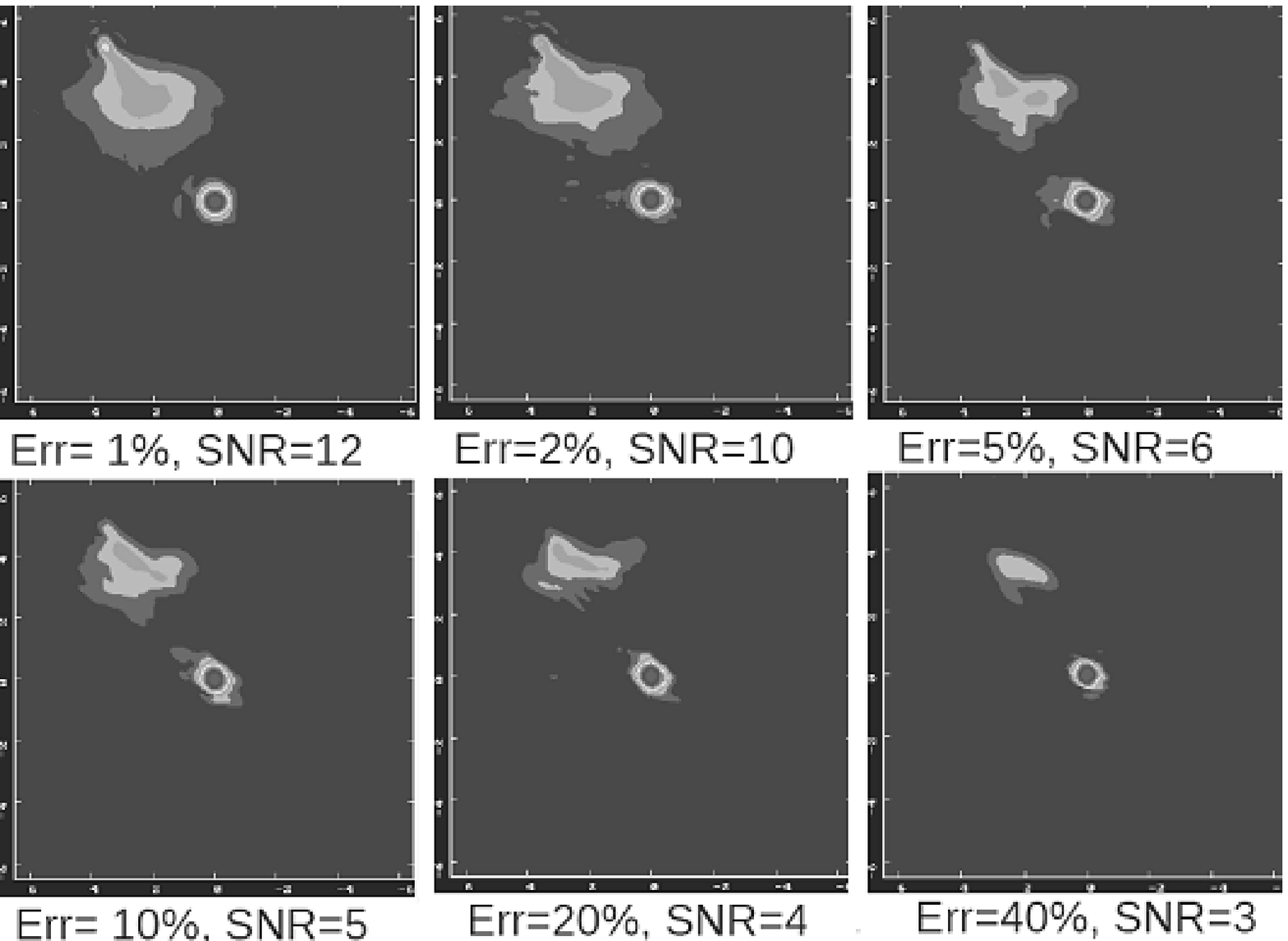}
 \caption{
Image reconstruction quality vs visibility magnitude errors for
source-4 }

\end{center}}
\end{figure}

The proposed technique was used for mapping seven selected AGNs
characterized by complicated structure of jets. We used VLBA data
from NRAO archive. Corresponding $UV$ coverages, intermediate GMEM
images and final images obtained using Fienup's phase retrieval
procedure are shown in Fig.11-17. These Figures contain all basic
information on both input data and images obtained. We can see,
that all intermediate images contain true information about
structure of jets what led to easy reconstruction of spectral
phase of sources on the final stage of reconstruction using
Fienup's algorithm. The images obtained are in good agreement with
ones published in literature what indicates high reliability of
our approach for VLBI imaging.

\begin{figure}[p]
{\begin{center}
 \includegraphics*[width=160mm]{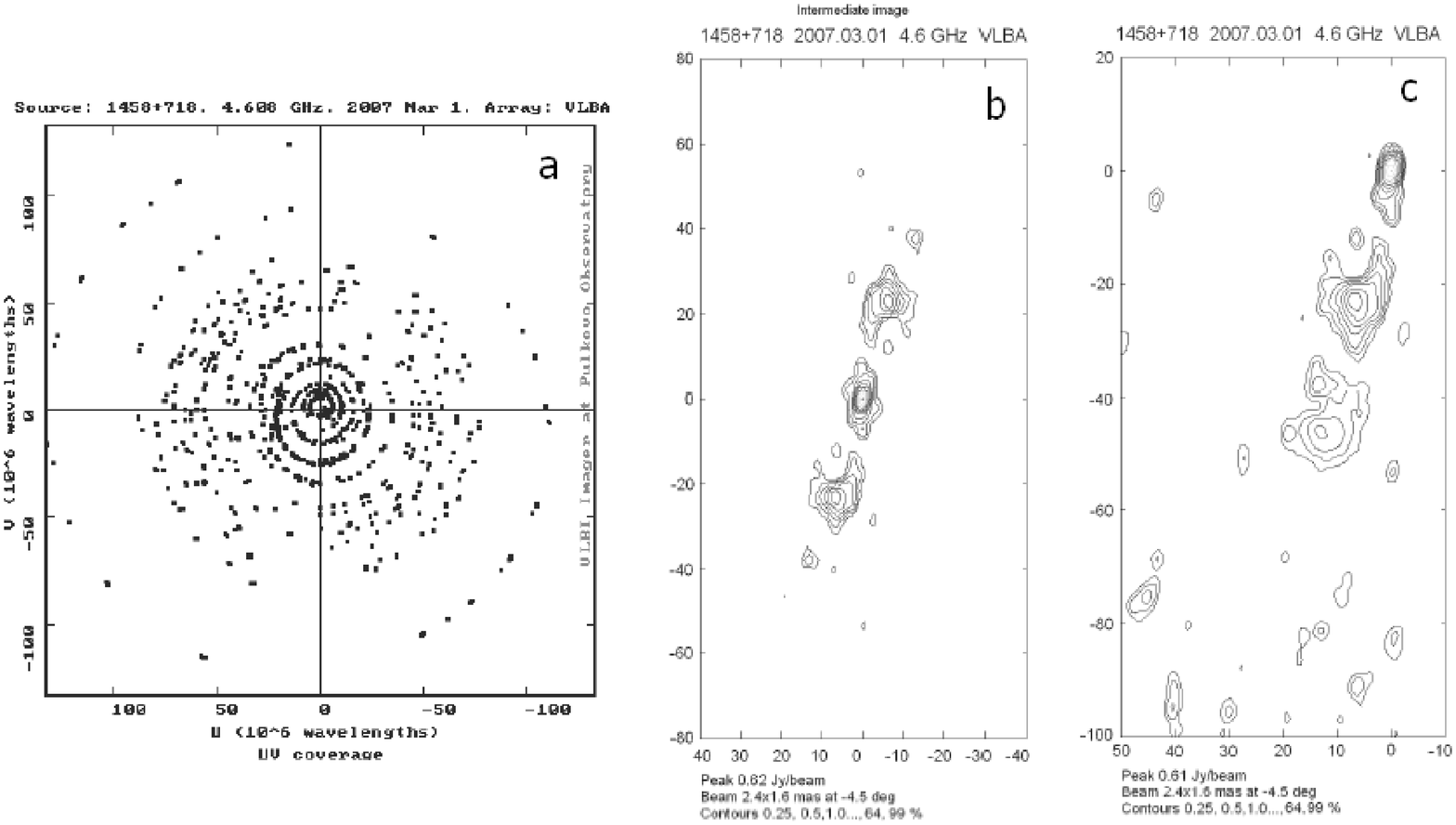}
 \caption{
"Phaseless" mapping of 1458+718. (a) $UV$ coverage, (b)
reconstructed intermediate GMEM image with zero-valued spectral
phase, (c) Fienup’s algorithm output image - solution of phase
retrieval problem}

\end{center}}
\end{figure}

\begin{figure}[p]
{\begin{center}
 \includegraphics*[width=160mm]{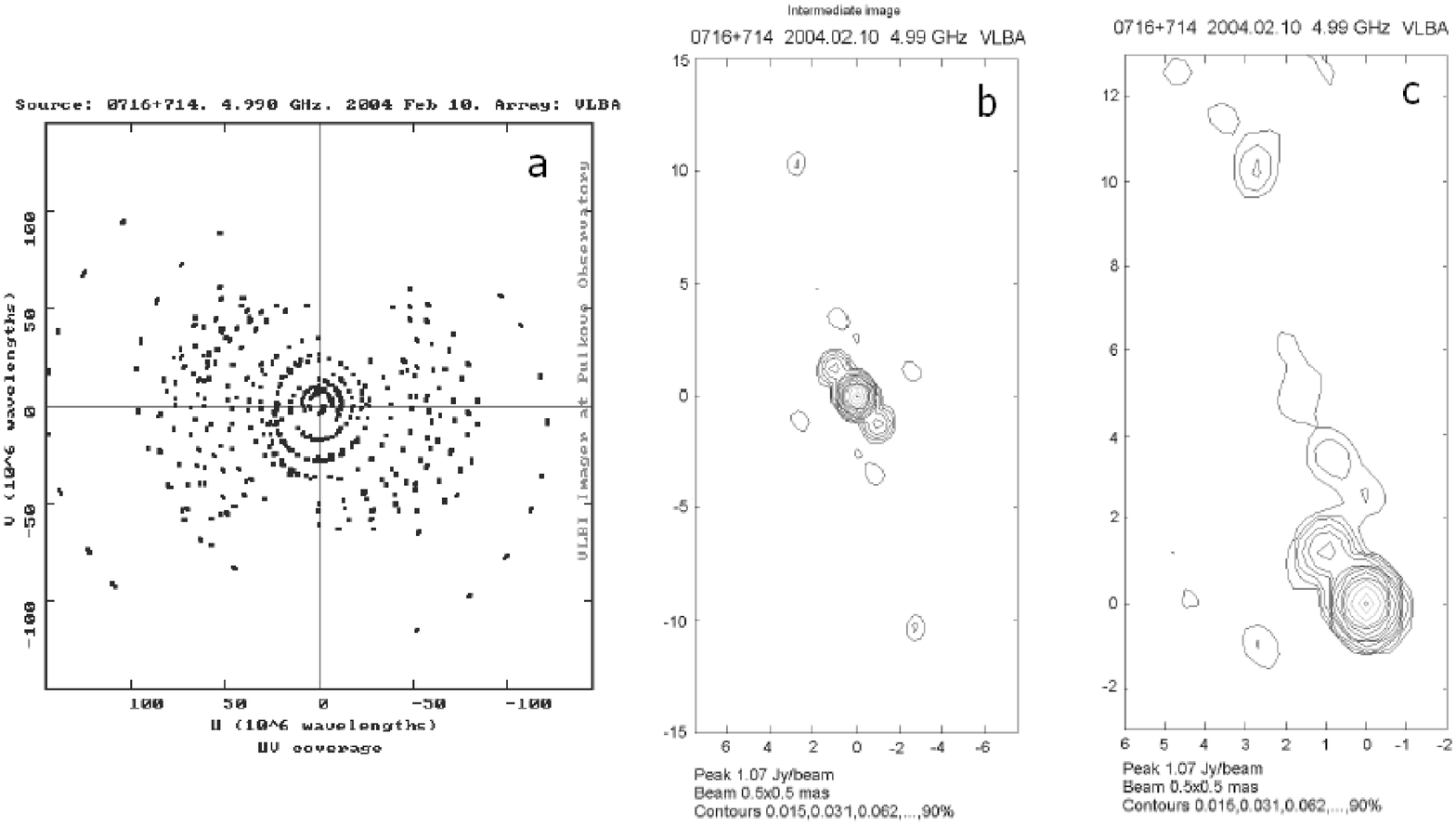}
 \caption{
"Phaseless" mapping of 0716+714. (a) $UV$ coverage, (b)
reconstructed intermediate GMEM image with zero-valued spectral
phase, (c) Fienup’s algorithm output image - solution of phase
retrieval problem}

\end{center}}
\end{figure}

\begin{figure}[p]
{\begin{center}
 \includegraphics*[width=160mm]{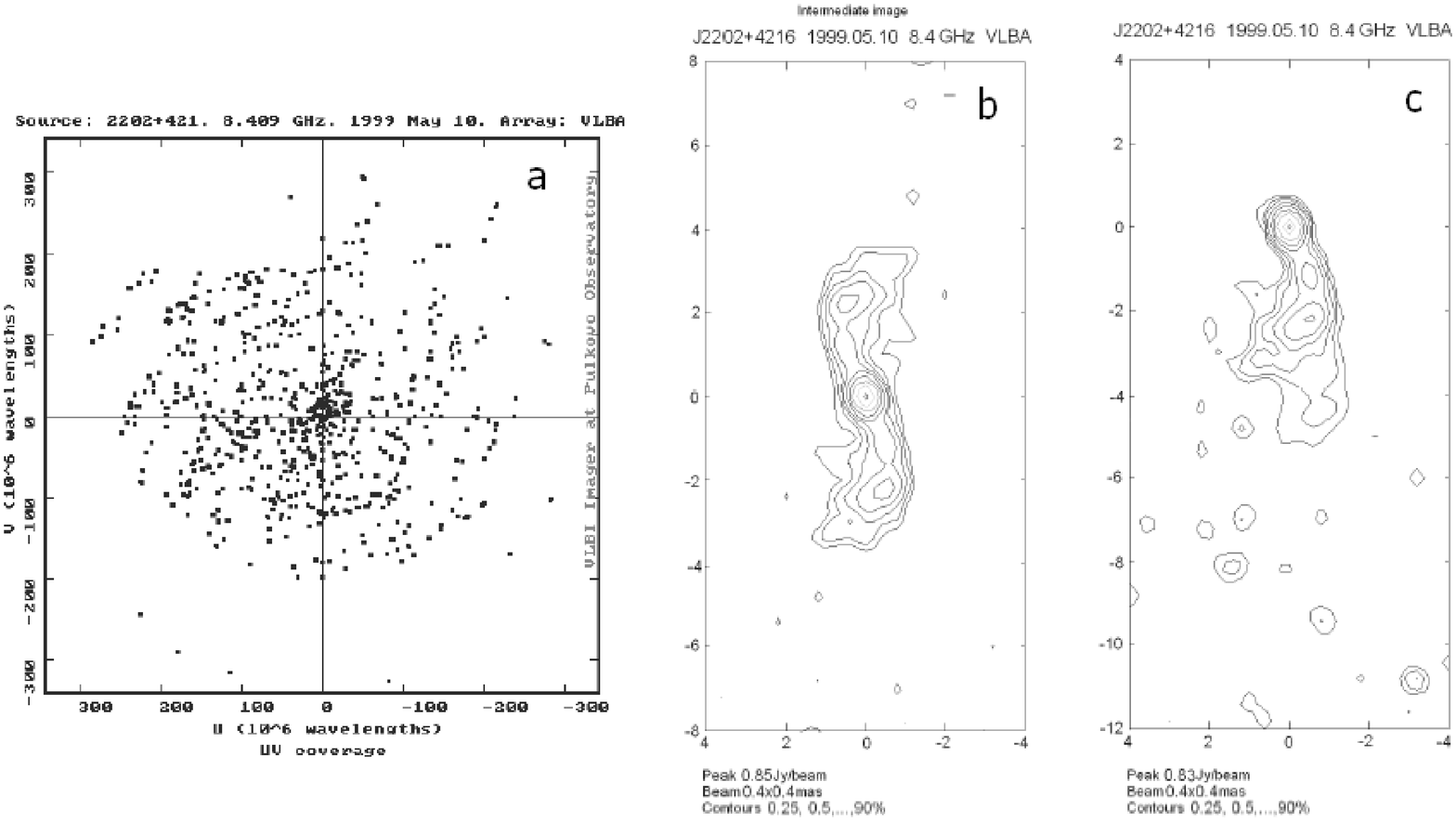}
 \caption{
"Phaseless" mapping of J2202+421. (a) $UV$ coverage, (b)
reconstructed intermediate GMEM image with zero-valued spectral
phase, (c) Fienup’s algorithm output image - solution of phase
retrieval problem}

\end{center}}
\end{figure}

\begin{figure}[p]
{\begin{center}
 \includegraphics*[width=150mm]{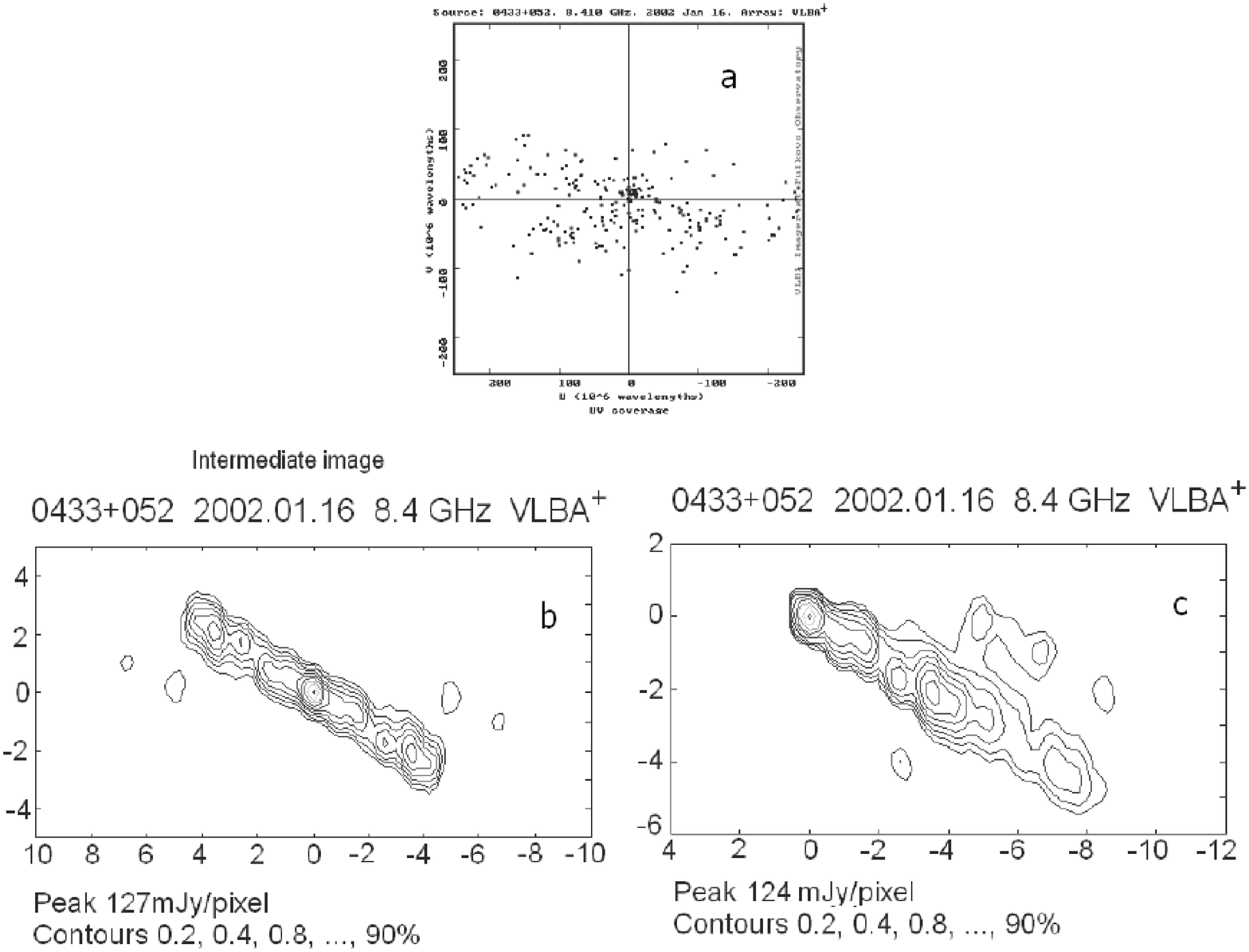}
 \caption{
"Phaseless" mapping of 0433+052. (a) $UV$ coverage, (b)
reconstructed intermediate GMEM image with zero-valued spectral
phase, (c) Fienup’s algorithm output image - solution of phase
retrieval problem}

\end{center}}
\end{figure}

\begin{figure}[p]
{\begin{center}
 \includegraphics*[width=140mm]{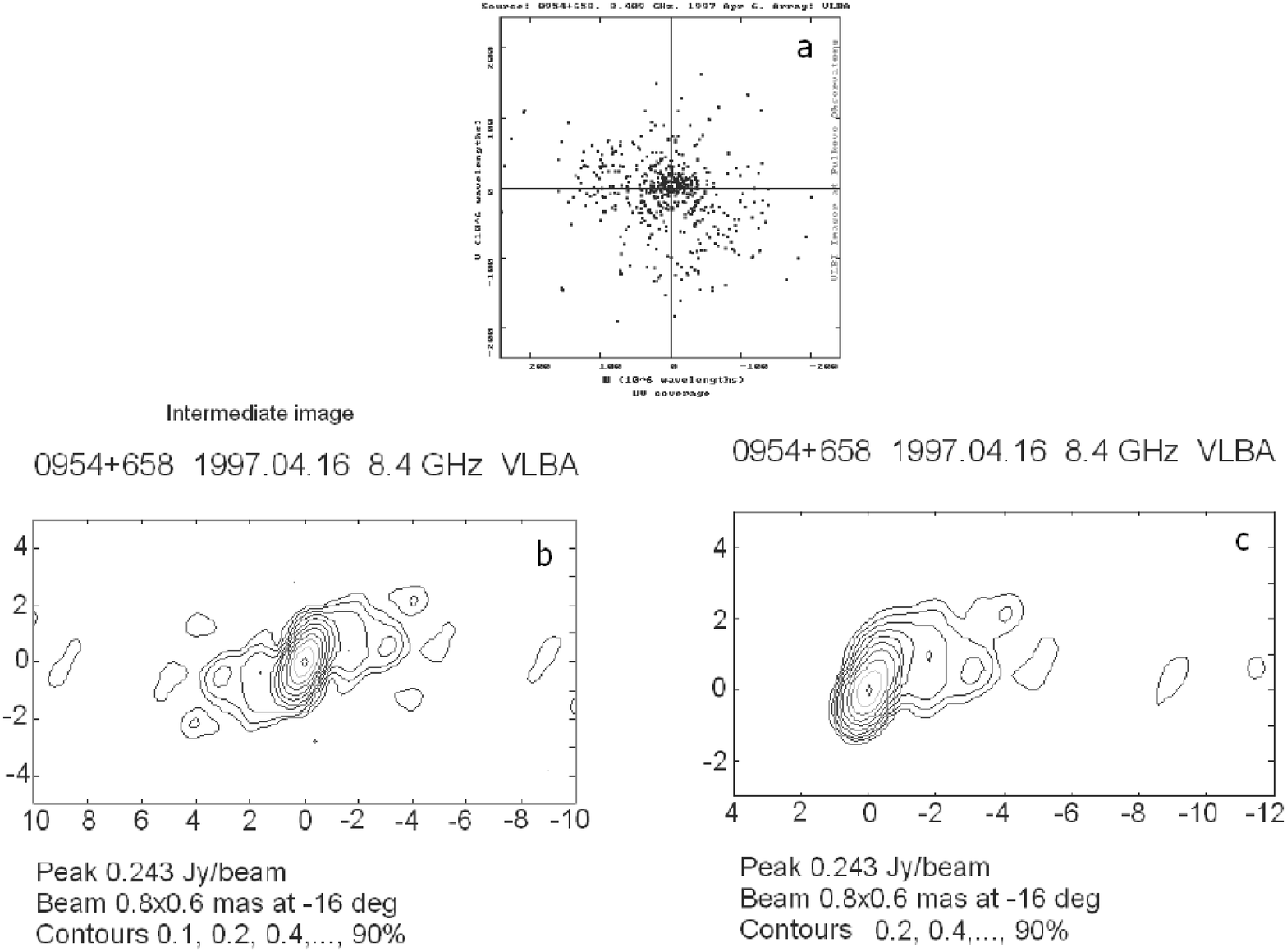}
 \caption{
"Phaseless" mapping of 0954+658. (a) $UV$ coverage, (b)
reconstructed intermediate GMEM image with zero-valued spectral
phase, (c) Fienup’s algorithm output image - solution of phase
retrieval problem}

\end{center}}
\end{figure}

\begin{figure}[p]
{\begin{center}
 \includegraphics*[width=140mm]{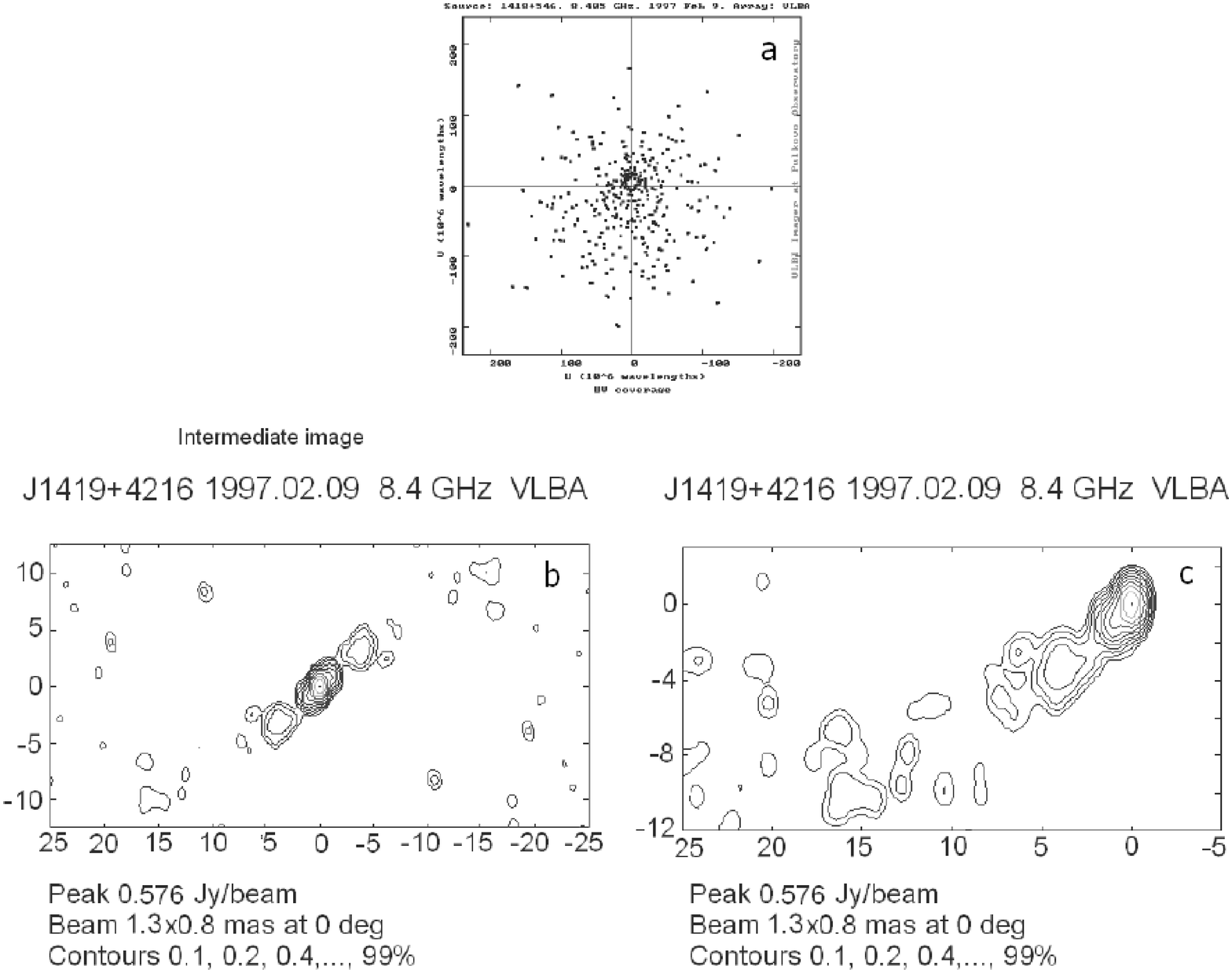}
 \caption{
"Phaseless" mapping of J1419+4216. (a) $UV$ coverage, (b)
reconstructed intermediate GMEM image with zero-valued spectral
phase, (c) Fienup’s algorithm output image - solution of phase
retrieval problem}

\end{center}}
\end{figure}

\begin{figure}[p]
{\begin{center}
 \includegraphics*[width=155mm]{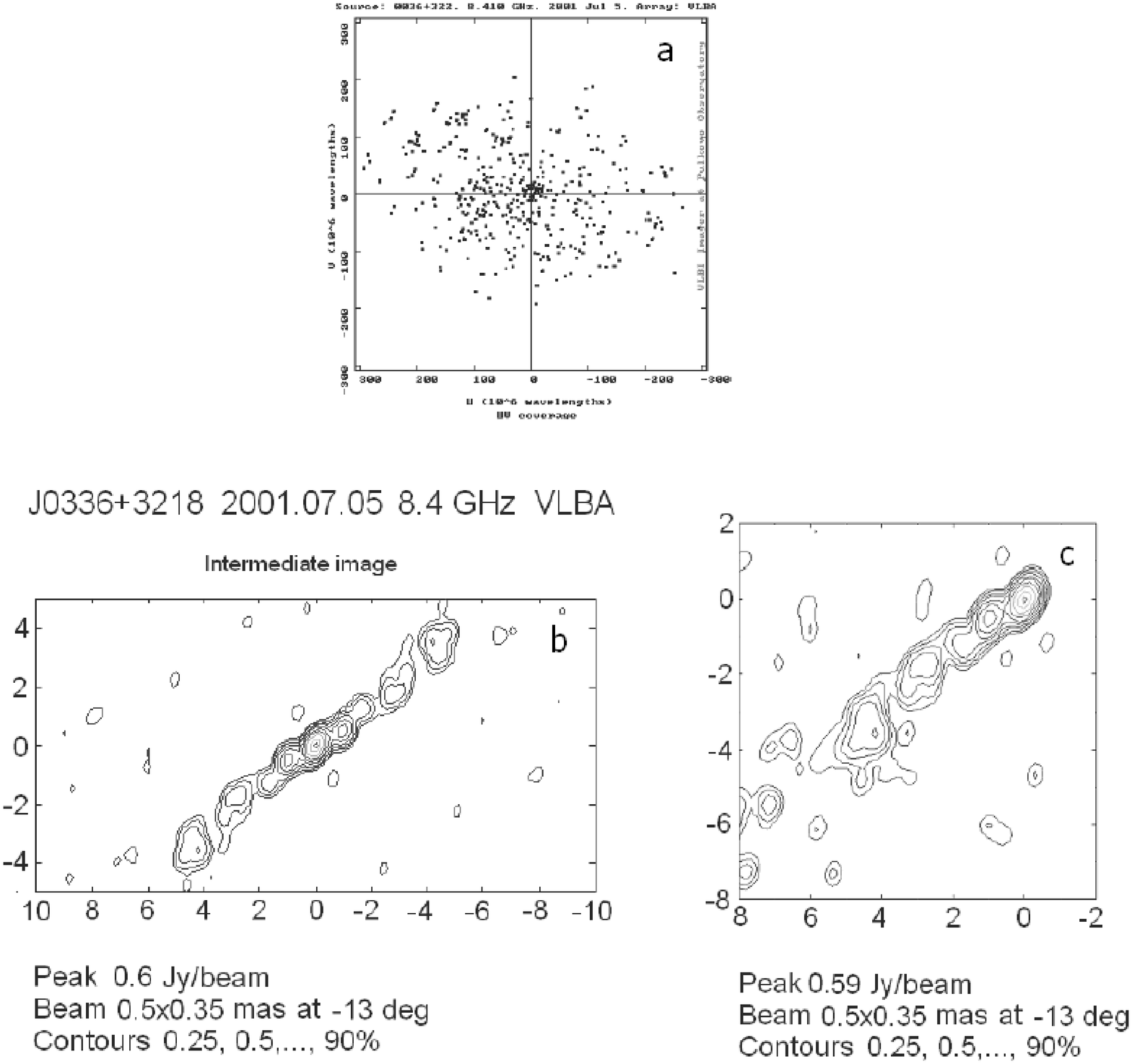}
 \caption{
"Phaseless" mapping of J0336+3218. (a) $UV$coverage, (b)
reconstructed intermediate GMEM image with zero-valued spectral
phase, (c) Fienup’s algorithm output image - solution of phase
retrieval problem}

\end{center}}
\end{figure}

\section{Modelling of "phaseless" mapping for "RadioAstron" mission}

Here we consider the problem of mapping with ultra-high angular
resolution using a space--ground radio interferometer with a space
antenna in a high orbit, whose apogee height exceeds the radius of
the Earth by a factor of tens.  "RadioAstron" mission (Kardashev
1997) represents such a system. Its basic parameters
(http://~www.~asc.~rssi.~ru/radioastron/description/\\orbit\_eng.htm)
are presented in Fig.18.

In this case, a multi-element interferometer essentially
degenerates into a two-element interferometer (Fig.19(a)). The
degeneracy of the close-phase relations prevents the use of
standard methods for hybrid mapping and self-calibration for the
correct reconstruction of images. We suggest that our "phaseless"
mapping approach based on methods for the reconstruction of images
from only visibility magnitudes (in the complete absence of phase
information) can be used in order to achieve the highest
resolution of the system.

In Fig.19(b-c) results of modelling of the "phaseless" mapping of
compact radio source having week jet component is shown.
Nevertheless the $UV$ coverage of the system is very poor (see
Fig.19(a)) the main featured of the source structure were
recovered.

\begin{figure}[p]
{\begin{center}
 \includegraphics*[width=160mm]{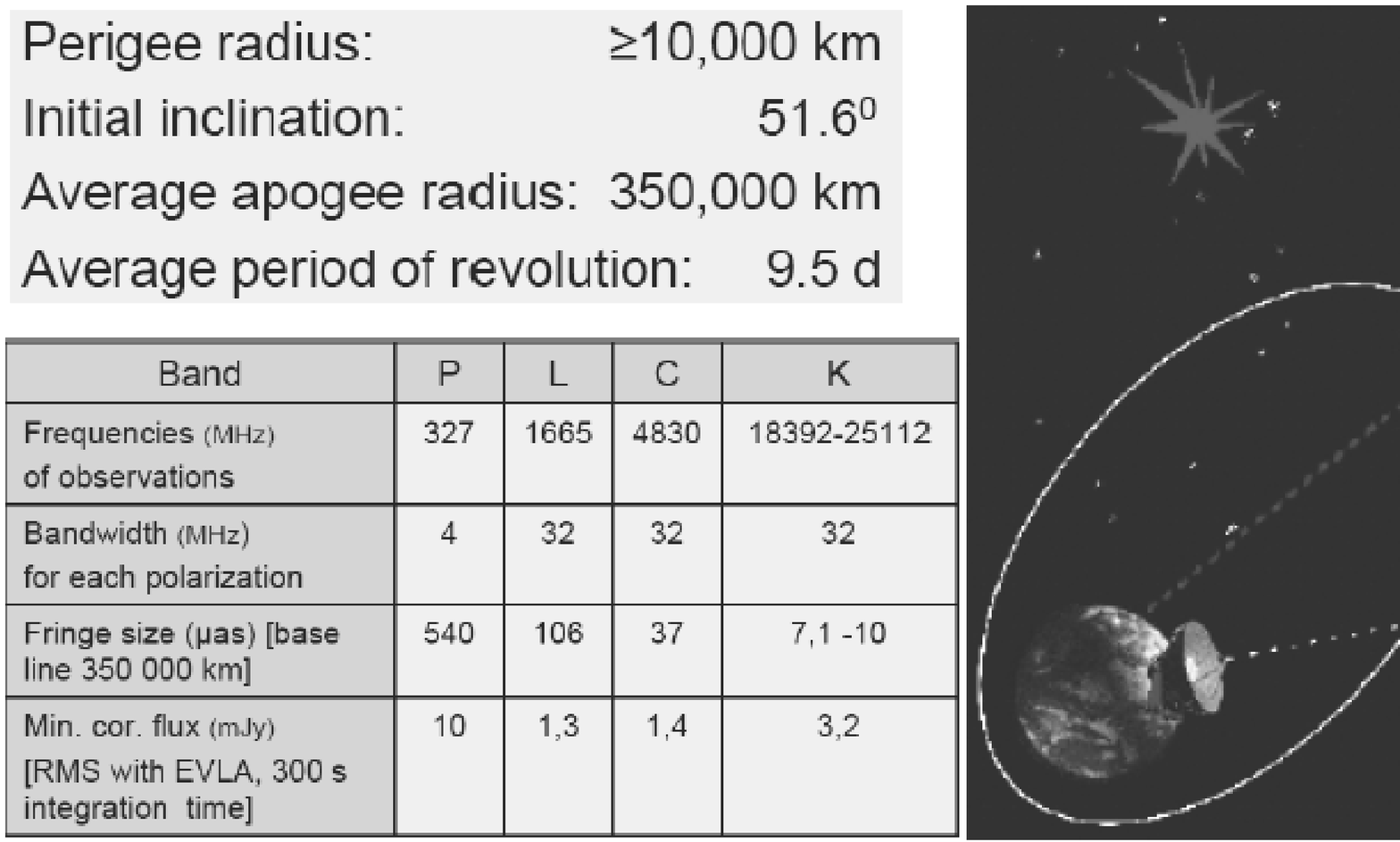}
 \caption{
Parameters of Space-Ground Radio Interferometer "RadioAstron"}

\end{center}}
\end{figure}

\begin{figure}[p]
{\begin{center}
 \includegraphics*[width=160mm]{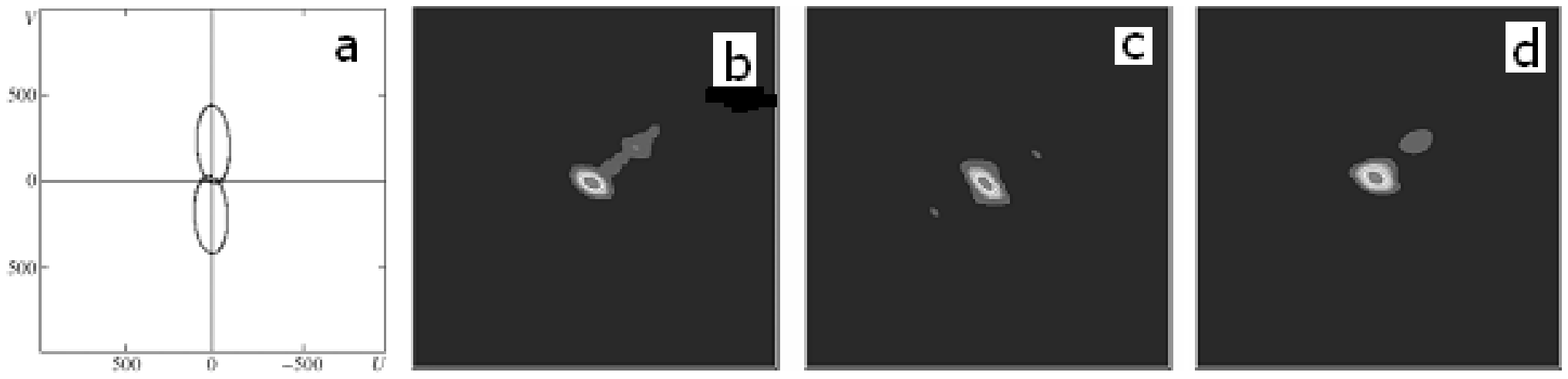}
 \caption{
Modelling of "phaseless"  imaging in "RadioAstron" system: (a)
$UV$ coverage (the scales along the axes are in units of $10^8$
wavelengths), (b) model source, (c) reconstructed intermediate
GMEM image with zero-valued spectral phase, (d) Fienup’s algorithm
output image - solution of phase retrieval problem}

\end{center}}
\end{figure}

\section{Conclusions}

\hskip 0.6cm 1.AGNs with structure "bright core + weak jet" are
quite suitable objects for "phaseless" VLBI mapping  based on
standard reconstruction procedures (MEM or CLEAN for deconvolution
and Fienup’s iterative algorithm for phase retrieval).

\medskip

2."Phaseless" mapping is urgent for "RadioAstron" mission because
of degenerate closure phases. In order to realize the highest
resolution ensured by high-orbit space-ground radio
interferometer, it is necessary to use methods for retrieving
phase information directly from visibility magnitude measurements.

\subsection*{Acknowledgments}

{\small This work was supported by the "Nonstationary Phenomena in
Objects of the Universe" Program of the Presidium of the Russian
Academy of Sciences and by the "Multiwavelength Astrophysical
Research" grant no. NSh--16245.2012.2 from the President of the
Russian Federation.}

\bigskip{\bf REFERENCES}\bigskip {\small

A.T. Bajkova, Astron. Astrophys. Trans. 1, 313 (1992)

A.T. Bajkova, Astronomy Letters, 30, 218 (2004)

A.T. Bajkova, Astronomy Reports, 49, 973 (2005)

Yu.M. Bruck and L.G. Sodin, Optics Comm., 30, 304 (1979)

T.J. Cornwell, R. Braun, and D.S. Briggs, in {\it Synthesis
Imaging in Radio Astronomy II. A Collection of Lectures from the
Sixth NRAO/NMIMT Synthesis Imaging Summer School}, Ed. by
G.~B.~Taylor, C.~L.~Carilli, and R.~A.~Perley (Astron. Soc. Pac.,
San Francisco, 1999); Astron. Soc. Pac. Conf. Ser., 180, 151
(1999).

J.C. Dainty, and M.A. Fiddy, Optica Acta, 31, 325 (1984)

J.R. Fienup, Opt.Lett., 3, 27 (1978)

J.R. Fienup, Applied Optics, 21, 2758 (1982)

J.R. Fienup,T.R. Crimmins,and W. Holsztynski, JOSA, 72, 610 (1982)

J.R. Fienup, JOSA, 73, 1421 (1983)

J.R. Fienup et al., General Dynamics Distinguished Lecture Series,
February 9, 2006

J.R. Fienup et al.,  Workshop on X-ray Science at the Femtosecond
to Attosecond Frontier, UCLA, May 19, 2009

B.R. Frieden and A.T. Bajkova, Appl. Opt., 33, 219 (1994)

R.Gerchberg and W.O. Saxton, Optik, 35, 237 (1972)

M.H. Hayes,IEEE Trans.Acoust.,Speech,Signal Process., 30, 140
(1982)

N.S. Kardashev, Exp. Astron., 7, 329 (1997)

Ya.I. Khurgin and V.P.Yakovlev, {\it Finite Functions in Physics
and Engineering} (Nauka, Moscow,1971) [in Russian ].

A.V. Oppenheim and J.S. Lim, Proc.IEEE, 69, 529 (1981)

J.L.C.Sanz and T.S.Huang, JOSA, 73, 1442 (1983)

{\it Image Recovery. Theory and Application}, Ed.by H. Stark
(Academic, Orlando,1987; Mir,Moscow, 1992)

}

\end{document}